\documentclass[12pt]{emulateapj}

\shorttitle{High angular resolution SZ effect in Cl~J0152-1357}
\shortauthors{Massardi et al.}

\begin{document}


\title{High angular resolution observation of the Sunyaev-Zel'dovich effect
 in the massive $z\approx0.83$ cluster Cl~J0152-1357.}


\author{M. Massardi}
\affil{INAF - Osservatorio Astronomico di Padova, vicolo dell'Osservatorio 5,I-35122, Italy}
\email{marcella.massardi@oapd.inaf.it}
\author{R. D. Ekers}
\affil{Australia Telescope National Facility, CSIRO Astronomy and Space Science, PO Box 76, Epping NSW 1710, Australia}
\author{S. C. Ellis}
\affil{Sydney Institute for Astronomy, School of Physics, University of Sydney, NSW 2006, Australia}
\and
\author{B. Maughan\altaffilmark{1}}
\affil{HH Wills Physics Laboratory, University of Bristol,Tyndall Ave, Bristol BS81TL,UK}
\altaffiltext{1}{Harvard-Smithsonian Center for Astrophysics, 60 Garden St, Cambridge, MA02140, USA.}


\begin{abstract}
X-ray observations of  galaxy clusters at high redshift ($z\gtrsim 0.5$) indicate that they are more morphologically complex and less virialized than those at low-redshift.  We present the first subarcmin resolution at 18 GHz observations of the Sunyaev-Zel'dovich (SZ) effect for Cl~J0152-1357 using the Australia Telescope Compact Array. Cl~J0152-1357 is a massive cluster at redshift $z=0.83$ and has a complex structure including several merging subclumps which have been studied at optical, X-ray, and radio wavelengths. Our high-resolution observations indicate a clear displacement of the maximum SZ effect from the peak of X-ray emission for the most massive sub-clump. This result shows that the cluster gas within the cluster substructures is not virialised in Cl~J0152-1357 and we suggest that it is still recovering from a recent merger event. A similar offset of the SZ effect has been recently seen in the `bullet cluster' by Malu et al.

This non-equilibrium situation implies that  high resolution observations are necessary to investigate galaxy cluster evolution, and to extract cosmological constraints from a comparison of the SZ effect and X-ray signals.

\end{abstract}



\section{Introduction}

The Sunyaev-Zel'dovich (SZ) effect is the spectral distortion of the Cosmic Microwave Background (CMB) due to inverse Compton scattering of CMB photons off a cloud of electrons (Birkinshaw 1999). The increase of the photon energy implies a decrease of the CMB brightness at frequencies below 218 GHz and an increase at higher frequencies. The amplitude of the distortion depends only on the properties of the electron cloud: in the case of thermal electrons it is proportional to the integral of the electron pressure (i.e. $\Delta T_{SZ}\propto n_eT_e$) along the line of sight, and is independent of distance. Galaxy clusters are massive structures permeated by hot dense ionized gas that preserves the same baryonic fraction of the epoch of the cluster virialization, so they are the ideal targets for SZ effect observations and to infer pieces of information about cosmology.

The X-ray luminosity from the hot cluster gas has a different dependence on gas properties (i.e. $L_X\propto n_e^2T_e^{1/2}$) and it is dependent on the cluster redshift. For virialized clusters, where the distribution of density and temperature are easy to model with regular forms (e.g. isothermal and spherically symmetric) the different distance dependence, provides a powerful method to extract redshift and cosmological parameters by combining the two signals (Cooray 1999, Carlstrom, Holder \& Reese 2002, Molnar, Birkinshaw \& Mushotzky 2002).

The comparison of the two signals is also a particularly effective tool to estimate the gas mass but the different dependence on density makes X-ray more susceptible to the gas clumping factor $C=\langle n_e^2\rangle^{1/2}/\langle n_e\rangle$, that may be $\gg 1$ at high redshift. The SZ effect signal also is a better tracer of less dense hotter region.

Current theories of structure formation predict that clusters form hierarchically via merger of smaller structures (Borgani \& Guzzo 2001). X-ray observations (Jeltema et al. 2005, Maughan et al. 2008) of high redshift ($z\gtrsim 0.5$) clusters revealed that they are more morphologically complex, less virialized and dynamically more active than low-redshift clusters. Studies of $z>0.8$ clusters show clumpy and elongated structures that suggest that they are close to the epoch of cluster formation (Rosati et al. 2004).

The study of high redshift cluster SZ effect can provide constraints on the theories of cluster formation and evolution. However, for these objects the temperature and density distributions might be so complex that the comparison of X-ray and SZ effect signals is seriously compromised. If this is the case the information about cosmology or cluster evolution will be misleading.

In this Letter we present the first subarcminute resolution 18 GHz observations of SZ effect with the Australia Telescope Compact Array using the new wide bandwidth 4 GHz correlator (Ferris \& Wilson 2002) for a high redshift cluster. The properties of the cluster and a summary of its existing observations in radio, optical and X-ray are described in \S \ref{sec:cl0152}. In \S \ref{sec:atdata} we present our new observations. Their comparison with other bands and the implications of our results are discussed in \S \ref{sec:discussion}. To estimate distances we have adopted a flat $\Lambda$CDM cosmology with $\Lambda=0.73$ and ${\rm H}_0 = 71\, {\rm km\, s^{-1}\, Mpc^{-1}}$.

\section{Cl~J0152-1357}\label{sec:cl0152}
\object{Cl~J0152-1357} is one of the most massive ($M_{tot}=1.1\times10^{15}M\odot$) galaxy clusters known at high redshift ($z\approx0.83$). It was discovered independently in the Wide Angle ROSAT Pointed Survey (Scharf et al. 1997, Ebeling et al. 2000), in the ROSAT Deep Cluster Survey (Della Ceca et al. 2000), and catalogued also in the Bright SHARC survey (Romer et al. 2000).

X-ray images from BeppoSax (Della Ceca et al. 2000), Chandra and XMM (Maughan et al., 2003 and 2006, Huo et al. 2004), optical images (Ellis \& Jones 2004, Girardi et al. 2005, Demarco et al. 2005, Kodama et al.2005, Jorgensen et al. 2005, Burbidge, Gutierrez \& Arp 2006), IR images (Maughan et al. 2006, Marcillac et al. 2007), and  weak lensing (Jee et al. 2005) observations all show a complex structure that appears to be far from virialized.

The cluster contains two main sub-clumps, 95 arcsec apart (corresponding to $\sim723$ kpc) and several other smaller structures that may be merging. The two main subclumps are aligned in the NE-SW direction (and for this reason we will dub them `NE' and `SW'). X-ray temperatures (Maughan et al. 2003) indicate that they have quite similar total masses, but the gas in NE one may be a little more massive. There is a suggestion in the X-ray of the presence of an excess of emission in the region in between them, possibly indicating that the two clumps are interacting and merging. X-ray observations indicate that the NE subclump is more extended, and slightly more massive than the SW subclump.

It has already been noted that the peaks of the galaxy mass distribution do not coincide with the X-ray emission peaks (Hou et al. 2003), or the weak lensing observations (Jee et al. 2005). However, it is a third smaller subclump in a SE region (see Girardi et al. 2005) that may be the key to the explanation of these observations.

Burbidge, Gutierrez \& Arp (2006) found a QSO at the same redshift 14 arcmin away ($\sim$6.45 Mpc) towards North-East, indicating that this cluster is part of a much larger scale structure. A similar indication comes also from the Subaru wide field imaging of the cluster field by Kodama et al. (2005).

Redshift distributions of the galaxies (Demarco et al. 2005) in the cluster region show similar distribution for all clumps at $z \approx 0.83$ but there is another group of galaxies at $z \approx 0.64$ along the same line of sight to the cluster.

The cluster SZ effect was detected by Joy et al. (2001) with the Berkley-Illinois-Maryland Association (BIMA) millimeter interferometer at 28.5 GHz  with a $-2.4\pm0.32$ mJy beam$^{-1}$ peak and a resolution of $151"\times 88"$ with a peak. This detection did not have high enough angular resolution to observe the effects of internal structure in this cluster. For the Northern subclump of this cluster Zemcov et al. (2007) may have detected the SZ increment from the SCUBA data at 350 GHz.

\section{Radio frequency observations}\label{sec:atdata}
\begin{figure}
\epsscale{0.8}
\plotone{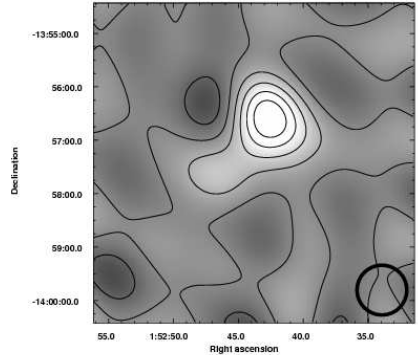}
\caption{SZ detection in the region of the NE subclump. Contours are multiple of 2$\sigma$. The 35 arcsec FWHM beam is represented by the solid ellipse on the bottom left.}\label{fig:sz}
\end{figure}
In 2005 we had a 24 hour observation with the `old' ATCA in its most compact configuration, Hybrid H75, with baselines from 30 to 75\ m. In this experiment we used two adjacent 128 MHz bands at 18.5 GHz. The primary beam FWHM was $\simeq 2.4$ arcmin with angular resolution of $33"\times 33"$. We made a mosaic of two pointings with centers on the X-ray position of each of the sub-clusters (NE 01h52m44s.18, -13$^\circ$57'15''.84, SW: 01h52m39s.89, -13$^\circ$58'27''.48).
We detected at a low significance level (signal to noise ratio equal to 3.5) a negative peak in the region of the NE subclump, but $\sim 35$ arcsec displaced towards the North-West from the position of the X-ray peak. A 1.5$\sigma$ dip at the position of the second main subclump suggests that the SZ peak of SW subclump maybe in the region of its X-ray expected position.

After the upgrade of the correlator of the Australia Telescope Compact Array to the new Compact Array Broadband Backend (CABB) digital correlator in July 2009 (Ferris \& Wilson 2002) we performed a new observing run with the same H75 configuration, but with 2$\times$2 GHz adjacent bands in two orthogonal polarizations (i.e. a 8 GHz total bandwidth) between 16 and 20 GHz. The angular resolution is 35"$\times$35" (FWHM). We observed 27 hours on the position of the X-ray NE peak. The negative peak is $-118.6\pm 11.53\ \mu$Jy (i.e. signal to noise ratio 10.3) in the position 01h52m42s.661 -13$^\circ$56'36''.40 (see Fig. \ref{fig:sz}). The single pointing was chosen to optimise our sensitivity to check the possible displaced SZ effect signal seen in the previous observations of the NE subclump at the sacrifice of good sensitivity in the SW subclump region. Future deeper observations will be carried out to investigate the SW subclump.
\begin{figure}
\epsscale{0.9}
\plotone{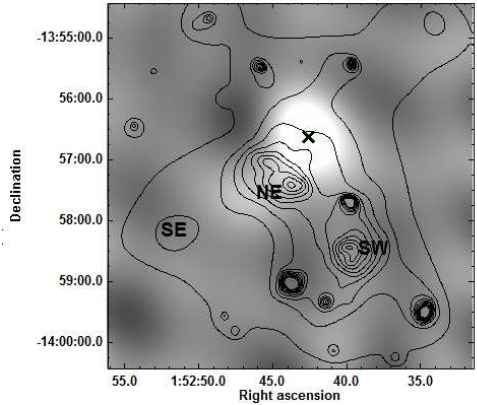}
\caption{XMM-Newton contours (Maughan et al. 2006) superimposed on 18 GHz ATCA image for
Cl~J0152-1357. The cross indicates the position of the SZ peak. Flags indicate the peaks of the subclumps mentioned in the text.}\label{fig:comp_x}
\end{figure}
\begin{figure}
\epsscale{0.9}
\plotone{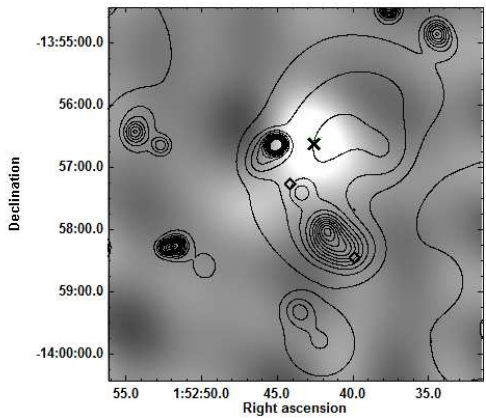}
\caption{Galaxy number density contours (Maughan et al. 2006) superimposed on 18 GHz ATCA image for Cl~J0152-1357. The comparison indicates a galaxy overdensity in the direction of the maximum SZ effect peak. The cross indicates the position of the SZ peak. Diamonds indicates the X-ray positions of the subclumps mentioned in the text}\label{fig:comp_gal}
\end{figure}

Fig. \ref{fig:comp_x} shows the ATCA SZ effect image with X-ray contours by Maughan et al. (2006) superimposed. Fig. \ref{fig:comp_gal} shows the same ATCA image with the galaxy density contours superimposed (Maughan et al. 2006).

We also observed for 3 hours at 1.4 GHz. The image has an rms of 4.8 mJy/beam and doesn't show any AGNs or diffuse emission within 2.5 arcmin of the cluster centre. As pointed out in the literature (e.g. for the cluster `bullet' cluster 1ES 0657-56 see Liang et al. 2000 and references therein) merging clusters often show non-thermal radio relics or haloes due to synchrotron emission in the region of the merging front. Since we observe no trace of any of these structures we have no supporting evidence for strong merging in this system. Two radio AGNs were detected far from the cluster region and they align exactly with two NVSS sources, as do two X-ray point sources outside the cluster central region which also have NVSS counterpart (too faint to affect our 18 GHz observations), showing that there is no position offset between X-ray and radio images.

\section{Discussion}\label{sec:discussion}

\subsection{The comparison with other bands}

The angular resolution of both the X-ray and radio images is sufficient to resolve the offset structures so the displacement can not be a resolution effect. We also know from the agreement in radio and X-ray positions of the two AGN just outside the cluster that the two coordinate systems are aligned to at least arcsecond accuracy.

However, from Fig. \ref{fig:comp_x} it is clear that the SZ effect  from the NE subclump is shifted $\sim45$ arcsec ($\sim$342 kpc) towards the North with respect to the X-ray peak of emission and is about 30 arcsec West of the main galaxy oncentration (see Fig. \ref{fig:comp_gal}).
A similar offset of the SZ effect has been recently seen in the `bullet cluster' by Malu et al. (ApJL submitted).

The cluster is very dynamically complex, and the northern subcluster is at the intersection between two merger axes and may have undergone mergers along both axes recently

The position of the maximum SZ effect ($\Delta T \propto  n_e T_e$) should trace the position of the gas pressure peak of this sub-clump but the X-ray (free-free) emission ($L_X \propto n_e^2 T_e^{1/2}$),  which also agrees with the weak-lensing-based mass distribution (Jee et al. 2005), is displaced 45 arcsec to the South-East so we need to explain this difference.

It is possible that the quality of X-ray data in the region of the SZ peak is relatively low, so that the temperature ($\sim4.4\pm1$ keV in Maughan et al. 2006) appears a bit lower than the peak of X-ray signal, triggering the displacement. Deeper X-ray observations will be considered to discard this possibility, but we can assume that at a first approximation we are observing a genuine displacement of the two signals due to the gas property distributions.

The group of galaxies at $z \sim 0.6$ could affect the X-ray emission more than the SZ effect, because it is closer, but it is a relatively low density group with a larger scale N-S distribution not clearly related to the observed X-ray emission. We thus consider it unlikely that the foreground structure is responsible for the observed offset.

In a relaxed structure the shape of the gravitational potential close to the centre of the mass distribution will determine both the X-ray and the SZ effect distribution. Hence, we expect that the signal from the SZ effect and the free-free X-ray emission from the intracluster medium to peak in the same position. This means that we must have different distributions of electron density ($n_e$) and temperature ($T_e$) in this region.

Since the weak lensing peak seems to be better align with X-ray rather than with SZ the displacement is most likely to be caused by an increased gas clumping factor at the position of the X-ray peak or a temperature excess in the position of the SZ peak.

The suggestion of a merging front between the NE and SW clumps (Maughan et al. 2003) might indicate a shocked higher density region, but the suggested location of this merging front is too far away to cause the 45 arcsec displacement. Furthermore if these two sub-clumps are merging for the first time it is hard to see how it could already strongly effect the region near the NE clump and far in front of the proposed merging front.

\begin{figure}
\epsscale{0.8}
\plotone{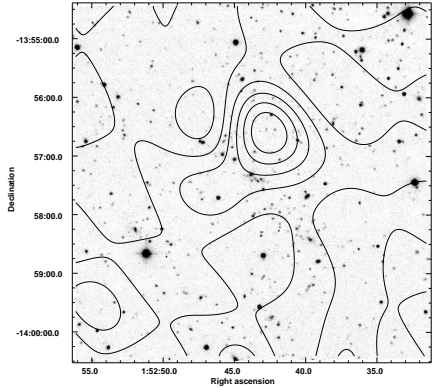}
\caption{SZ effect contours as in Fig. \ref{fig:sz} overlapped to the K-band IRIS2 image.}\label{fig:opt_sz}
\end{figure}

The galaxy distribution (Demarco et al. 2005) in the region of the SZ peak seems to indicate the presence of few galaxies (see Fig. \ref{fig:opt_sz}). No spectra are currently available for them, but photometric analysis will be used to estimate their redshift. Those galaxies could also be part of the cluster, representing yet another merger component along that SE-NW axis.

A more likely scenario is that this displacement is a post merger effect due to the passage of a smaller clump (e.g. the one to the SE) through the denser NE clump. Maughan et al. (2006) suggested that the NE-SW merger happened more recently than a merger with the SE group, relying on the fact that the projected separation of the structures along the NE-SW direction is much smaller than the separation between NE and SW subclumps, and because the X-ray morphology appears more disrupted along this NE-SW axis in the Northern subcluster.
However, the passage of the SE subclump through the NE one would destroy the equilibrium in the NE clump and could generate regions of shocked gas causing excess X-ray emission on one side (including also an elongation of the NE X-ray peak towards North, perpendicularly to the crossing direction). It would also heat the gas in the potential of the NE clump further enhancing the SZ effect relative to the X-ray emission in this region.

It is also plausible that there is a mixture of cooler gas associated with the cluster and much hotter gas [including possibly very hot $\gtrsim20$ keV gas that is hard to detect with XMM or Chandra as shown for the `bullet' cluster by Markevitch et al. (2002)] associated with the merger, projected along the line of sight. Depending on the densities of these components, the X-ray and SZ effect would respond very differently to the two components, with the hotter components going undetected in the current X-ray data. This, along with the relatively poor X-ray statistics in this region could explain the lack of evidence for enhanced pressure at the SZE peak in the XMM data.

Deeper high resolution SZ (including the SW peak) and X-ray observations might help finding a conclusive description for the dynamics of this cluster.

\subsection{The impact on cosmology}
A clear consequence of our finding is that any cosmological conclusion based on the assumption that the SZ effect and the X-ray emission result from the same relaxed gas distribution could be wrong for clusters at high redshift like Cl~J0152-1357.
Low resolution SZ effect data do not allow  the proper density and temperature distribution to be disentangled, with the risk that if the peak of the signal is blindly associated with the X-ray peak an error in the estimation of cosmological parameters is introduced.

The cosmological parameters are usually estimated by comparing the angular diameter distance $D_A$ as estimated by the X-ray and SZ effect. In order to determine the distance it is necessary to properly model the structure of the intracluster medium, to connect the quantities determied among integrations along the cluster line of sight with the cluster appearence on the plane of the sky: typically isothermal and spherical symmetric density distributions are assumed to extract $D_A$ and cosmological parameters. The failures of the assumptions for several reasons, including mergers, asphericity, clumpiness and nonisothermality induces errors on the cosmological parameters which increase with redshift: numerical simulation (Molnar, Birkinshaw \& Mushotzky 2002, Yoshikawa, Itoh, and Suto 1998, Inagaki, Suginohara, \& Suto 1995, Roettiger, Stone, \& Mushotzky 1997, Mohr et al. 1999) have demonstrated that at $z\sim 1$ the cumulative effect of isothermality and asphericity, almost negligible at low redshift but can introduce up to a 20\% error in the estimation of the angular diameter distance and that the assumption of isothermality for merging subclusters overestimates $D_A$ by a 15-20\%. The error introduced by this offset is likely to be much larger.

The error is larger the farther from virialization, hence clusters at higher redshift which exhibit less virialized structures than local ones could be more seriously affected. This also has implications on the studies of cluster evolution, because estimates of barionic mass and intracluster medium properties might be wrong.

In conclusion we stress the importance that only virialized clusters can be of use for deriving cosmological parameters. Furthermore, wherever X-ray observations indicate that the structure of the cluster is not yet relaxed the high resolution SZ effect data is a powerful tool for the analysis of the cluster properties. In combination with X-ray observations of free-free emission it may provide a powerful tool to study the effects of shocks and other phenomena which will be important for our understanding of the evolution of galaxy clusters.

\acknowledgments
Partial financial support for this research has been provided to MM by the Italian ASI/INAF Agreement I/072/09/0 for the Planck LFI Activity of Phase E2 and I/016/07/0 `COFIS'.
SCE is funded by ARC FF grant 0776384 through the University of Sydney.

We thank the staff at the Australia Telescope Compact Array site, Narrabri (NSW), for the valuable support they provide in running the telescope. The Australia Telescope Compact Array is part of the Australia Telescope which is funded by the Commonwealth of Australia for operation as a National Facility managed by CSIRO.

This research has made use of the NASA/IPAC Extragalactic Database (NED) which is operated by the Jet Propulsion Laboratory, California Institute of Technology, under contract with the National Aeronautics and Space Administration.



\clearpage



\end{document}